\documentclass[11pt]{article}
\usepackage[final]{acl}

\usepackage{booktabs}
\usepackage{multirow}
\usepackage{graphicx}
\usepackage{subcaption}
\usepackage{float}
\usepackage{amsmath}
\usepackage{amssymb}
\usepackage{xcolor}
\usepackage{placeins}
\usepackage{times}
\usepackage{latexsym}
\usepackage[T1]{fontenc}
\usepackage[utf8]{inputenc}
\usepackage{microtype}
\usepackage{inconsolata}

\usepackage{tikz}
\usetikzlibrary{arrows.meta, backgrounds}

\title{RetrievalRouter: Joint Modality and Architecture Selection for Document Retrieval}

\author{
  \textbf{Emre Kuru\textsuperscript{1}},
  \textbf{Mehmet Onur Keskin\textsuperscript{2}},
  \textbf{Reza Farahbakhsh\textsuperscript{1}},
  \textbf{Noel Crespi\textsuperscript{1}}
  \\
  \textsuperscript{1}SAMOVAR, Télécom SudParis, Institut Polytechnique de Paris, Palaiseau, France \\
  \textsuperscript{2}Özyeğin University, Istanbul, Türkiye \\
  \small{\texttt{emre.kuru@telecom-sudparis.eu}}
}

\begin{document}
\maketitle

\begin{abstract}

Document retrieval increasingly supports high-stakes information access in finance, healthcare, and law. Modern retrieval pipelines vary both in modality (text or multimodal) and in retrieval architecture (dense or late-interaction). These choices impose a hard compromise: the most effective pipelines are too slow and expensive to run at scale, while the fastest fail to retrieve evidence from complex documents. Practitioners must therefore choose between missed evidence and unusable latency, with no principled basis for adapting that choice at the query level. We show that this compromise is unnecessary. Not every query requires the same pipeline. Across benchmarks spanning financial and scientific corpora, no static pipeline dominates. We introduce RetrievalRouter, a lightweight query-aware router that learns, from the query text alone, which retrieval pipeline best fits each query. A single tunable parameter exposes the full accuracy--latency frontier, and for every static baseline, RetrievalRouter offers an operating point that is simultaneously more accurate and faster. Against the best static baseline, RetrievalRouter is 2.5\% more accurate and $12.4\times$ faster. Furthermore, compared with prior adaptive strategy selection methods, RetrievalRouter achieves significantly higher nDCG@5 across accuracy-oriented settings, while matching or numerically outperforming them on both nDCG@5 and latency in latency-oriented settings. Our code: \url{https://github.com/emrekuruu/retrieval-router}

\end{abstract}
\section{Introduction}
\label{sec:introduction}

Document retrieval underpins information access in high-stakes domains such as finance \cite{li2025fingear}, healthcare \cite{xia2024rule}, and law \cite{gao2024enhancing}. In these settings, retrieving the right evidence is essential, as missing or irrelevant documents can lead to unsupported critical decisions.

Modern document retrieval pipelines vary along two design axes. The first is modality: text-based retrievers operate on text extracted from documents \cite{lin2022retrieval}, while multimodal retrievers \cite{ma2024unifying, faysse2024colpali} operate directly on rendered page images. The second is architecture: dense retrievers \cite{karpukhin2020dense} compress each document into a single embedding for fast retrieval, while late-interaction architectures \cite{khattab2020colbert} preserve fine-grained per-token embeddings at higher cost. These choices are not free. Accuracy and efficiency move in opposite directions across the space: configurations that handle complex documents most reliably are also the slowest, while configurations cheap enough for production traffic fail silently on those same documents~\cite{kuru2026evaluatingmodernragtextual}. Without a principled way to predict which pipeline a given query needs, practitioners must commit to a single configuration at design time and pay the cost on every query.

We show that this compromise is unnecessary. Not every query requires the same pipeline. Across 11 retrieval benchmarks spanning financial, scientific, and open-domain corpora, we find that each pipeline's failures are query-dependent and asymmetric. Along the modality axis, text-based pipelines degrade sharply on queries against visually complex documents, while multimodal pipelines fall behind on queries that require nuanced textual understanding. Along the architecture axis, dense retrievers match late-interaction performance on the majority of queries at a fraction of the cost, but late-interaction architectures remain essential on the subset of queries where dense matching fails.

To address this gap, we introduce \textbf{RetrievalRouter}, a lightweight router that, given only the query text, predicts both the modality and the architecture of the retrieval pipeline to apply. To our knowledge, RetrievalRouter is the first system to route jointly across these two axes for a single underlying corpus. By directing each query to the cheapest pipeline that can answer it, the router reserves expensive configurations for queries that genuinely require them. The resulting system attains the accuracy of the strongest static pipeline while incurring an average latency near the cheapest. In summary, our main contributions are:

\begin{itemize}
    \item A systematic empirical analysis of modern retrieval pipelines along two axes, modality and architecture, evaluated across 11 benchmarks, and characterizing the query-level strengths and failure modes of each configuration.
    
    \item RetrievalRouter, the first query-aware router that jointly selects retrieval modality and architecture per query. It dominates every static configuration on the accuracy--latency frontier, achieving 2.5\% higher nDCG@5 than the strongest static baseline while being $12.4\times$ faster. Compared with prior adaptive strategy selection methods, it achieves significantly higher nDCG@5 in accuracy-oriented settings, while numerically outperforming them in both effectiveness and latency in latency-oriented settings.

    \item A query-level pipeline-selection benchmark, releasing per-query best-pipeline labels across more than 80{,}000 queries to support further research on adaptive retrieval.
\end{itemize}

\section{Related Work}
\label{sec:related}

This work builds on two lines of prior research: the static retrieval pipelines whose design space we route over, and the adaptive retrieval strategies that explore query-time decision-making.

Early document retrieval systems applied dense retrievers to extracted document text \cite{lin2022retrieval, karpukhin2020dense}. To recover the document structure lost in extraction, subsequent work introduced layout-aware chunking that segmented documents into semantic regions such as headers and tables \cite{yepes2024financial}. The text-extraction step itself remains a fundamental ceiling: \citet{zhang2025ocr} shows that extraction errors propagate through the pipeline causing systematic failures on layout-sensitive queries that dense text embeddings cannot recover.

Multimodal retrievers emerged as a response, encoding rendered page images directly and avoiding text extraction. MuRAG \cite{chen2022murag} introduced joint image-text retrieval through a multimodal memory; later work embedded both modalities into a unified representation space and enabled dense retrieval over raw visual inputs \cite{ma2024unifying, riedler2024beyond}. ColPali \cite{faysse2024colpali} extended the ColBERT late-interaction paradigm to visual patches, achieving state-of-the-art accuracy on visually complex benchmarks. However, visual encoding has its own ceiling: VTCBench \cite{zhao2025vtcbench} shows that vision-language models degrade substantially on tasks requiring long-range textual reasoning, falling behind text-only retrievers in such settings.

A growing body of work relaxes the assumption that a fixed retrieval pipeline serves every query. Adaptive approaches differ in \emph{what they adapt}. One line decides \emph{whether} to retrieve or \emph{how deeply}, conditioning on query difficulty, uncertainty, or learned policies \cite{asai2024self, tang2025mba}, while keeping the retriever itself fixed. A second line routes across heterogeneous knowledge bases or distinct source modalities \cite{peng2025learning, jiang2025qa}, where each corpus serves a different semantic role.

Closer to our setting, a recent line of work routes across different retrievers over a single corpus. \citet{arabzadeh2021predicting} train a query-only classifier to choose between sparse and dense retrieval using hard strategy labels derived from retrieval success. LiteGator \cite{darmanto2025litegator} switches between sparse and dense retrieval per query under a latency budget. RouterRetriever \cite{lee2025routerretriever} routes among domain-specific LoRA experts within a single dense architecture. MoR \cite{kalra2025mor} ensembles sparse and dense retrievers with per-query trust weights computed from pre- and post-retrieval signals.
\section{Background}
\label{sec:landscape}

We evaluate seven retrieval pipelines spanning two modalities (text and multimodal) and four retrieval strategies (sparse, dense, late-interaction, and reranking). The five core pipelines define the design space; the two reranking variants represent the systems deployed in practice.

\paragraph{Text pipelines.} 
Sparse pipelines (BM25) apply term-based matching to extracted page text \cite{robertson2009probabilistic}. Unlike neural retrievers, BM25 requires no embedding generation, making it the lowest-latency pipeline in our design space.
Dense pipelines (TD) embed the same extracted page text into a single vector with a bi-encoder (Linq-Embed-Mistral \cite{LinqAIResearch2024}). Their learned representations capture semantic similarity beyond exact term overlap, improving retrieval when queries and relevant pages use different wording, but embedding generation makes TD slower than BM25. 
Late-interaction pipelines (TL) swap the bi-encoder for a late-interaction model (GTE-ModernColBERT \cite{GTE-ModernColBERT}) and preserve per-token embeddings, enabling finer-grained matching at higher cost. TL is the most accurate core text pipeline, but also the slowest. 
All three text pipelines inherit the limitations of text extraction: extraction errors and layout loss cap their accuracy on visually structured documents \cite{zhang2025ocr}. Because none of the text pipelines can perceive non-textual content, we caption charts, figures, and tables with a vision-language model (Gemini 3.0 Flash \cite{team2023gemini}) and concatenate the descriptions with the extracted text before indexing \cite{zhao2023retrieving}, so any remaining performance gap reflects architectural limits rather than missing content.

\paragraph{Multimodal pipelines.} Multimodal-Dense (MD) operates directly on full-page images without text extraction, encoding each page into a single visual embedding (Nomic-Embed-Multimodal \cite{nomicembedmultimodal2025}). It is the fastest neural pipeline overall and the natural multimodal counterpart to TD. Multimodal-Late (ML) keeps the visual input but represents each page as a bag of patch embeddings (ColNomic-Embed-Multimodal \cite{nomicembedmultimodal2025}) and applies patch-level late interaction at retrieval time. It is the most accurate pipeline, but also the slowest.

\paragraph{Reranking pipelines.} Reranking is the standard industry technique for deploying late-interaction at scale, recovering nearly all of its accuracy at a fraction of its latency \cite{nogueira2019passage}: a fast dense retriever fetches a small candidate set ($k{=}100$), and the late-interaction model is applied only to those candidates. We include both rerank variants because they reflect how late-interaction is actually deployed in practice and they constitute the strongest deployable static baselines our router must surpass. Text-Rerank (TR) combines Linq-Embed-Mistral with GTE-ModernColBERT, and Multimodal-Rerank (MR) combines Nomic-Embed-Multimodal with ColNomic-Embed-Multimodal.
\section{Proposed Approach}
\label{sec:method}

The seven retrieval pipelines evaluated in Section~\ref{sec:landscape} occupy different points on the accuracy--latency frontier, and no single pipeline dominates. Any fixed choice, therefore, forces one trade-off on every query: overspend on queries it can handle easily, or underperform on queries it handles poorly. We propose RetrievalRouter, a lightweight query-aware policy that resolves this by selecting a retrieval pipeline for each query. From the query text alone, it predicts which pipeline best fits the query.

\begin{figure*}[h]
\centering

\definecolor{cardblueBG}{HTML}{E8EEF8}
\definecolor{cardblueLabel}{HTML}{2E4F7E}
\definecolor{cardyellowBG}{HTML}{FAF4DC}
\definecolor{cardyellowLabel}{HTML}{7A6122}
\definecolor{cardpurpleBG}{HTML}{EFE8F5}
\definecolor{cardpurpleLabel}{HTML}{5C3A85}
\definecolor{barred}{HTML}{C84444}
\definecolor{doctext}{HTML}{3A3A3D}
\definecolor{doctextlight}{HTML}{D0D0D2}

\begin{tikzpicture}[
  font=\sffamily,
  card/.style={
    rounded corners=10pt, line width=0.4pt,
    minimum width=4.8cm, minimum height=5.8cm
  },
  query/.style={font=\small\itshape, align=center, text width=4.3cm, black!85},
  pipeline/.style={font=\bfseries\small, align=center},
  justify/.style={
    font=\footnotesize, align=center, black!60,
    text width=4.4cm
  },
  arr/.style={->, >={Stealth[length=2.5mm, width=2mm]},
    line width=0.8pt, black!40},
]

\def\cAx{0}
\begin{scope}[on background layer]
  \node[card, fill=cardblueBG, draw=cardblueLabel!25] at (\cAx, 0) {};
\end{scope}

\node[pipeline, text=cardblueLabel] at (\cAx, 2.45) {Multimodal Dense};

\begin{scope}[shift={(\cAx-0.825, 1.2)}]
  \fill[barred] (0.00, 0) rectangle (0.25, 0.45);
  \fill[barred] (0.35, 0) rectangle (0.60, 0.70);
  \fill[barred] (0.70, 0) rectangle (0.95, 0.30);
  \fill[barred] (1.05, 0) rectangle (1.30, 0.55);
  \fill[barred] (1.40, 0) rectangle (1.65, 0.40);
  \draw[cardblueLabel!30, line width=0.5pt] (-0.05, 0) -- (1.70, 0);
\end{scope}

\node[query] at (\cAx, 0.4) {``What does the red bar chart showcase here?''};
\draw[arr] (\cAx, -0.15) -- (\cAx, -1.15);
\node[justify] at (\cAx, -1.85) {Text pipelines cannot see the color; multimodality is required to interpret the visual reference.};

\def\cBx{5.8}
\begin{scope}[on background layer]
  \node[card, fill=cardyellowBG, draw=cardyellowLabel!25] at (\cBx, 0) {};
\end{scope}

\node[pipeline, text=cardyellowLabel] at (\cBx, 2.45) {BM25};


\node[query] at (\cBx, 1.0) {``Who is the CEO?''};
\draw[arr] (\cBx, 0.6) -- (\cBx, -0.45);
\node[justify] at (\cBx, -1.5) {A direct keyword match identifies the answer; the cheapest pipeline suffices, heavier ones would add latency without improving accuracy.};

\def\cCx{11.6}
\begin{scope}[on background layer]
  \node[card, fill=cardpurpleBG, draw=cardpurpleLabel!25] at (\cCx, 0) {};
\end{scope}

\node[pipeline, text=cardpurpleLabel] at (\cCx, 2.45) {Text Late-Interaction};

\begin{scope}[shift={(\cCx-0.9, 1)}]
  \draw[cardpurpleLabel!35, fill=white, rounded corners=2pt, line width=0.4pt]
    (0, 0) rectangle (1.8, 1.0);
  \fill[doctext] (0.13, 0.87) rectangle (1.45, 0.94);
  \fill[doctext] (0.13, 0.76) rectangle (1.40, 0.83);
  \fill[doctextlight] (0.13, 0.63) rectangle (1.55, 0.69);
  \fill[doctextlight] (0.13, 0.30) rectangle (1.45, 0.36);
  \fill[doctext] (0.13, 0.17) rectangle (1.50, 0.24);
  \fill[doctext] (0.13, 0.06) rectangle (1.30, 0.13);
\end{scope}

\node[query] at (\cCx, 0.4) {``Did Q4 results match the Q1 projections?''};
\draw[arr] (\cCx, -0.15) -- (\cCx, -0.85);
\node[justify] at (\cCx, -1.85) {Evidence spans both ends of a long document; late-interaction preserves token-level structure that dense retrieval averages away.};

\end{tikzpicture}
\caption{Qualitative routing examples. The router sends queries to the cheapest pipeline capable of answering it.}
\label{fig:router-examples}
\end{figure*}
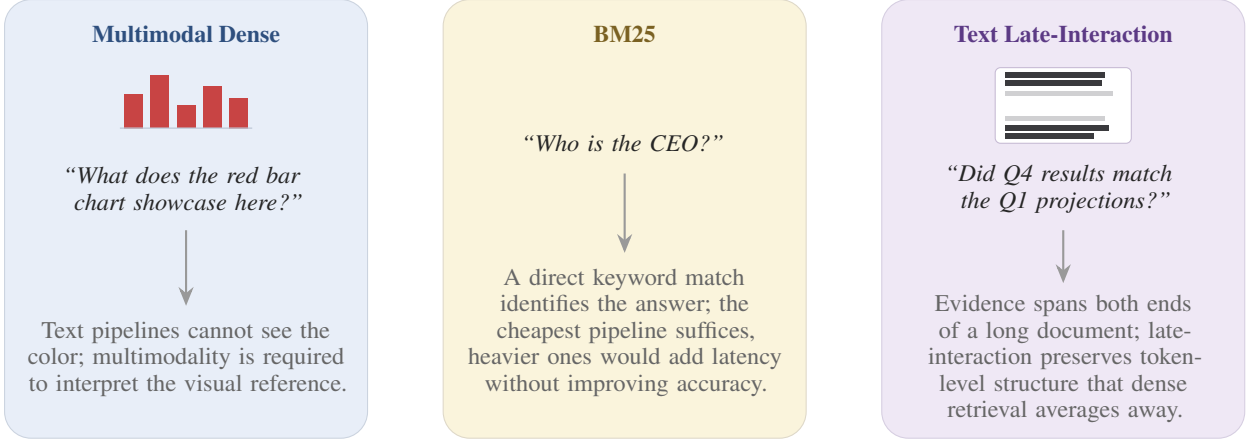

Figure~\ref{fig:router-examples} illustrates this on three representative examples. The first, a visual reference query, asks about a chart identified by its color, a property that text-only retrieval cannot perceive, so multimodal capabilities are required. The second, a textual factoid, can be resolved through direct term matching, so BM25 avoids unnecessary embedding generation. The third spans a long document with evidence at both ends; dense retrieval averages the document into a single vector, losing this distant structure, while late-interaction preserves the token-level vectors that capture evidence across the document. Each decision picks the cheapest pipeline that can answer the query, and the savings from easy queries make expensive configurations affordable for hard ones. No fixed-pipeline system has this property.

We use a single router over the five routing arms rather than splitting the decision into a modality router and an architecture router. Pipelines are not fully described by their modality and architecture alone; each has its own strengths and trade-offs, and the signals about which one fits a query live in the query's latent representation. A router with direct access to all five arms can learn these patterns; one that operates at a higher level does not see the full picture at once.

\subsection{Problem Formulation}

Let $\mathcal{P} = \{\text{BM25}, \text{TD}, \text{TL}, \text{TR}, \text{MD}, \text{ML}, \text{MR}\}$ denote the seven evaluated pipelines. In preliminary analyses, we found that the reranking pipelines matched or nearly matched the effectiveness of their pure late-interaction counterparts at substantially lower latency. Including both variants would introduce near-redundant actions and make the routing signal harder to learn. We therefore define the router action space as $\mathcal{A} = \{\text{BM25}, \text{TD}, \text{TR}, \text{MD}, \text{MR}\} \subset \mathcal{P}$, while retaining TL and ML as static baselines. 

The router is a parameterized policy $\pi_\theta(\cdot \mid q)$ that maps a query $q$ to a distribution over $\mathcal{A}$. At inference time, we route $q$ to the pipeline $\arg\max_{p_i \in \mathcal{A}} \pi_\theta(p_i \mid q)$. For each routing arm $p_i \in \mathcal{A}$ and query $q$, we define a per-query reward that combines accuracy and efficiency:

\begin{equation}
r_i(q) = (1 - \lambda) \cdot s_i(q) + \lambda \cdot \bigl(1 - \ell_i(q)\bigr)
\label{eq:reward}
\end{equation}

where $s_i(q) \in [0,1]$ is the nDCG@5 of routing arm $p_i$ on query $q$ and $\ell_i(q) \in [0,1]$ is its per-query normalized latency, $\ell_i(q) = \frac{L_i(q)}{\sum_{p_j \in \mathcal{A}} L_j(q)}$. Here, $L_i(q)$ is the raw wall-clock latency of routing arm $p_i$, and the denominator is the total latency across the five routing arms in $\mathcal{A}$. The efficiency score is then represented by $1 - \ell_i(q)$. The hyperparameter $\lambda \in [0,1]$ controls the accuracy--latency trade-off: $\lambda = 0$ yields a quality-only objective, and $\lambda \rightarrow 1$ favors the fastest routing arm. Sweeping $\lambda$ at training time traces out the accuracy--latency Pareto frontier.

\subsection{Router Architecture}

The router has two components: a query encoder $E_\phi$ and a lightweight decision head $H_\psi$. We use Qwen3-0.6B-Base \cite{team2025qwen3} as the encoder, with LoRA adapters \cite{hu2022lora} on the attention and feedforward projections; the base weights remain frozen during training. We mean-pool the encoder's final hidden states into a 1024-dimensional query representation $z_q$. The decision head is a single linear layer that maps $z_q$ to logits $\mathbf{s} \in \mathbb{R}^{|\mathcal{A}|}$ over the five routing arms, and the output distribution is a softmax over the logits, $\pi_\theta(p_i \mid q) = \exp(s_i) / \sum_{p_j \in \mathcal{A}} \exp(s_j)$.

\subsection{Training}

Training on hard labels (the single best pipeline per query) is unreliable for this problem. On any given query, multiple pipelines often retrieve the same documents and tie on nDCG@5, and forcing the router to pick one arbitrary winner injects label noise that the model has no way to resolve. We train instead against a soft target derived from the full per-query reward vector over $\mathcal{A}$.

\paragraph{Oracle labels.} For each training query $q$, we construct the reward vector over the five routing arms, $\mathbf{r}_{\mathcal{A}}(q) = [r_i(q)]_{p_i \in \mathcal{A}}$, using Equation~\ref{eq:reward}.

\paragraph{Soft targets.} The reward vector $\mathbf{r}_{\mathcal{A}}(q)$ is converted to a target probability distribution over the routing arms via softmax:

\begin{equation}
\tilde{p}_i(q) = \frac{\exp(r_i(q) / \tau)}{\sum_{p_j \in \mathcal{A}} \exp(r_j(q) / \tau)}
\label{eq:soft_target}
\end{equation}

where $\tau > 0$ is a temperature scaling parameter. We set $\tau = 0.1$. Because the rewards $r_i(q)$ are bounded in $[0,1]$ and concentrate in a narrow band when most pipelines succeed, a standard softmax ($\tau = 1$) yields near-uniform targets and dilutes the training signal, especially on easy queries where the only differentiator is latency. A small $\tau$ amplifies these gaps into a clear preference while exact ties remain exact ties. Queries with all-zero rewards are excluded from the gradient. At $\lambda=0$, this removes queries for which every arm fails; at $\lambda>0$, the latency term provides an efficiency signal, so these queries remain in training.

\paragraph{Objective.} The router minimizes the KL divergence between its predicted policy and the target:

\begin{equation}
\mathcal{L}(\theta) = \sum_{q \in \mathcal{D}_{\text{train}}} D_{\text{KL}}\bigl(\tilde{\mathbf{p}}(q) \, \big\| \, \pi_\theta(\cdot \mid q)\bigr).
\end{equation}

The router thus learns to approximate the per-query oracle decision at inference, without running any of the pipelines themselves. Implementation details are provided in Appendix~\ref{apx:experimental_details}.
\section{Experimental Setup}
\label{sec:setup}

We evaluate RetrievalRouter across 11 benchmarks spanning financial, scientific, and open-domain corpora. All retrieval pipelines, router training, and inference are run on the same NVIDIA H100 80GB GPU, ensuring latency measurements are directly comparable across systems.

\subsection{Datasets}
\label{subsec:datasets}

We use 11 datasets spanning text-heavy scientific papers, mixed-modality financial reports, and chart-dense slide decks, drawn from three benchmarks. Dataset statistics are reported in Appendix~\ref{apx:experimental_details}.

\begin{itemize}
    \item \textbf{REAL-MM-RAG} \cite{wasserman2025real}: \textit{FinReport} and \textit{FinSlides}. Real-world financial documents.
    \item \textbf{T2-RAGBench} \cite{strich2025t2ragbench}: \textit{FinQA}, \textit{ConvFinQA}, \textit{VQAonBD}, and \textit{TAT-DQA}. Hybrid reasoning over text and tables.
    \item \textbf{MMDocRAG} \cite{dong2025mmdocir}: \textit{ArxivQA}, \textit{Wiki-SS}, \textit{MP-DocVQA}, \textit{SciQAG}, and \textit{DUDE}. Long-context textual retrieval.
\end{itemize}

\subsection{Baselines}

We compare RetrievalRouter against the seven static pipelines from Section~\ref{sec:landscape} (BM25, TD, TL, TR, MD, ML, MR), each of which is run identically on every query. These establish the accuracy--latency frontier achievable without query-aware routing.

We also compare against the closest existing adaptive baseline to our setting, the query-level retrieval strategy-selection method introduced by \citet{arabzadeh2021predicting}. The original method trains a hard-label classifier to select a sparse retriever when it ranks a relevant document above a fixed threshold and otherwise escalate to a dense retriever. We extend this rule to our five-pipeline set by ordering the pipelines by their measured mean latency and assigning each training query to the cheapest pipeline that ranks a relevant page in the first position. Queries for which no pipeline succeeds are assigned to BM25. At inference time, a probability threshold controls whether a query remains on BM25 or is escalated to the classifier's highest-scoring neural pipeline. Sweeping this threshold produces the baseline’s accuracy--latency frontier. Here, budget denotes the fraction of queries routed away from BM25 to a neural pipeline. For comparison with RetrievalRouter, we report the baseline on the equivalent scale $\lambda=1-\mathrm{budget}$, so lower $\lambda$ corresponds to a larger escalation budget.

Finally, we report a seven-pipeline per-query Oracle as an upper bound. For each test query, the Oracle selects from the full evaluated set $\mathcal{P}$, including TL and ML, the pipeline that maximizes the reward from Equation~\ref{eq:reward}. For this upper bound, latency is normalized across all pipelines in $\mathcal{P}$ rather than the five routing arms in $\mathcal{A}$.

\subsection{Metrics}

\paragraph{Effectiveness.} We report nDCG@5, MRR@5, and Recall@5.

\paragraph{Latency.} End-to-end wall-clock time per query, including router inference, embedding generation, and vector search. We report both mean and P95 latencies to capture tail behavior.

\paragraph{Statistical testing.} We assess normality of paired-difference distributions using the D'Agostino--Pearson test, applying a paired $t$-test when normal and a two-sided Wilcoxon signed-rank test otherwise. We control family-wise error across planned comparisons using Holm correction. All statistical significance is reported at $p<0.001$.

\paragraph{Storage.} We report the disk footprint of the four neural vector indices maintained by RetrievalRouter. Rerank pipelines reuse the corresponding dense and late-interaction indices and therefore require no additional vector storage. BM25 uses a separate lexical index, which is not included in the vector-store comparison.
\section{Evaluation}
\label{sec:results}

\begin{figure*}[ht]
\centering
\includegraphics[width=0.94\linewidth]{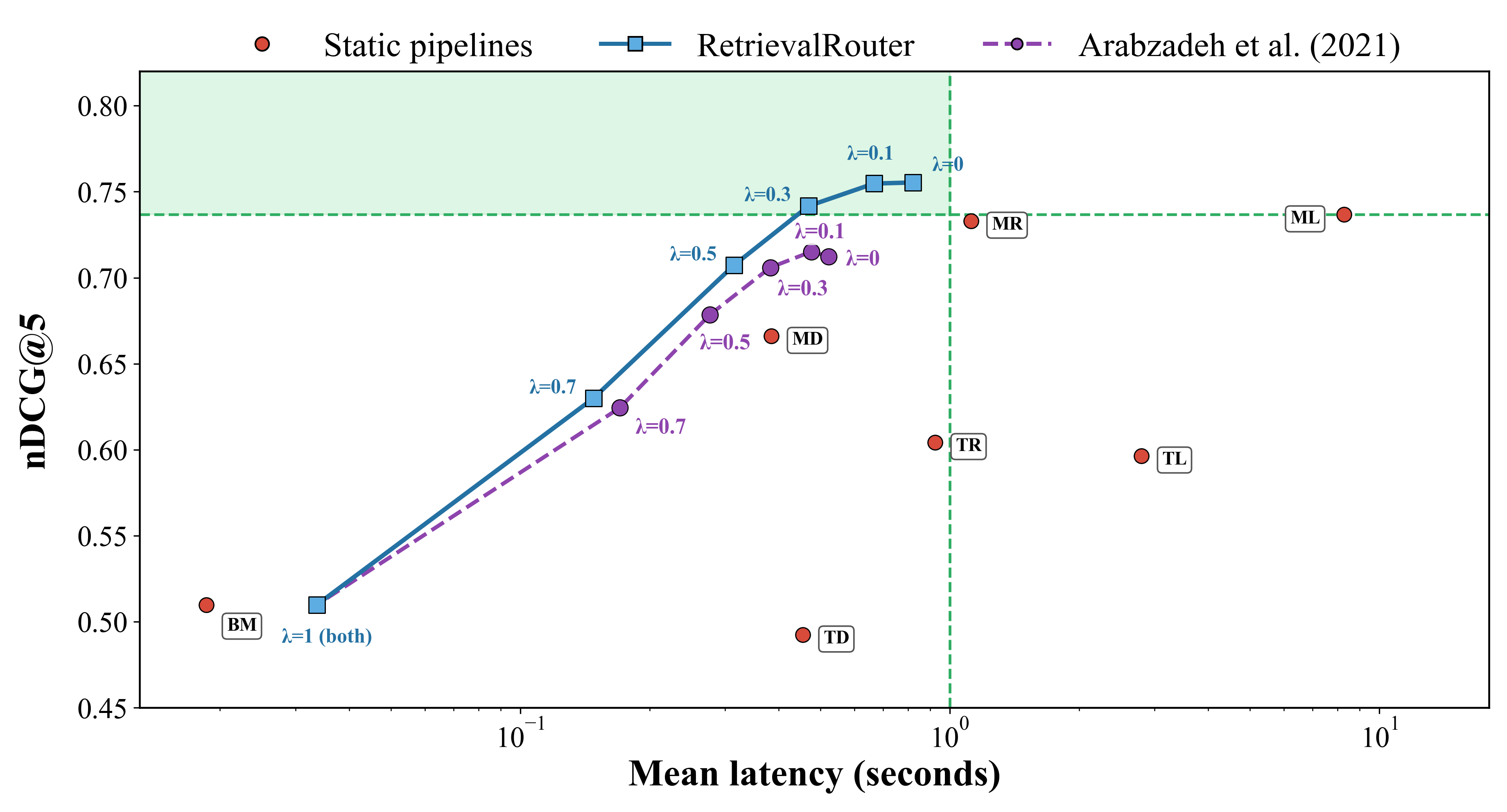}
\caption{Accuracy--efficiency operating points for all methods. The shaded area marks the wanted region: higher accuracy than the most accurate static pipeline with mean latency below 1s.}
\label{fig:pareto}
\end{figure*}


Table~\ref{tab:main_results} reports retrieval effectiveness and end-to-end latency for the seven static pipelines. Table~\ref{tab:adaptive_results} compares RetrievalRouter, Arabzadeh et al. (2021), and the Oracle at six values of $\lambda$. Figure~\ref{fig:pareto} plots the static pipelines and deployable adaptive methods in the accuracy--latency plane.
 
\paragraph{Static pipelines.} ML is the most effective static pipeline, reaching 0.737 nDCG@5, but is also the slowest at 8.283s per query. MR nearly preserves this effectiveness (0.733 nDCG) while reducing mean latency to 1.121s. MD provides a lower-cost middle ground at 0.666 nDCG and 0.385s. At the opposite extreme, BM25 is by far the fastest pipeline at 0.019s, but reaches only 0.510 nDCG. No static pipeline therefore combines the effectiveness of late interaction with the latency of lightweight retrieval.

\paragraph{RetrievalRouter dominates the neural static baselines on both axes.} Sweeping $\lambda$ along the accuracy--latency frontier yields operating points that dominate every neural static pipeline. At $\lambda=0.1$, RetrievalRouter reaches 0.755 nDCG@5 at 0.666s and dominates all four late-interaction pipelines. Relative to the multimodal variants, it improves effectiveness by 2.5\% over ML and 3.0\% over MR while being $12.4\times$ and $1.7\times$ faster, respectively. Relative to the text variants, it improves effectiveness by 26.5\% over TL and 24.9\% over TR while being $4.2\times$ and $1.4\times$ faster. At $\lambda=0.5$, RetrievalRouter reaches 0.707 nDCG@5 at 0.314s and dominates both dense pipelines, improving effectiveness by 6.2\% over MD and 43.6\% over TD while being $1.2\times$ and $1.4\times$ faster. Finally, at $\lambda=1$, RetrievalRouter selects BM25 for every query and reaches 0.034s total latency after policy inference. All gains are significant ($p<0.001$).

\paragraph{Router overhead.} Routing incurs 15 ms of overhead. This cost becomes meaningful in total latency only when a substantial share of queries is routed to BM25. At $\lambda=1$, where every query is routed to BM25, the overhead increases the latency from 0.019s to 0.034s; even then, it remains faster than every neural static pipeline.

\paragraph{RetrievalRouter outperforms prior adaptive routing.} In the accuracy-oriented settings ($\lambda=0$ through $0.5$), RetrievalRouter achieves significantly higher nDCG@5 than the strategy-selection baseline of \citet{arabzadeh2021predicting}, while the baseline remains significantly faster (all $p<0.001$). As $\lambda$ increases, RetrievalRouter's effectiveness gain narrows from 0.043 to 0.029 nDCG and its latency overhead falls from 0.299s to 0.038s. At the latency-oriented setting $\lambda=0.7$, RetrievalRouter numerically improves both effectiveness (0.630 vs. 0.624 nDCG) and mean latency (0.148s vs. 0.171s), although neither difference is significant (nDCG: $p=0.021$; latency: $p=0.015$). At $\lambda=1$, both methods select BM25 for every query and therefore converge to the same endpoint.

\paragraph{Oracle headroom.} The per-query oracle reaches 0.90 nDCG@5, leaving roughly 14 nDCG points of additional headroom available with perfect routing decisions. Although the oracle is not a deployable system, this gap demonstrates how far a query-aware router could plausibly go, given the current set of static pipelines.

\begin{table}[h]
\centering
\resizebox{\linewidth}{!}{%
\begin{tabular}{cccccc}
\toprule
\textbf{Model} & \textbf{nDCG} & \textbf{MRR} & \textbf{Recall} & \textbf{Latency} & \textbf{P95} \\
\midrule
\textbf{BM25} & 0.510 & 0.476 & 0.613 & \textbf{0.019} & \textbf{0.046} \\
\textbf{TD}   & 0.492 & 0.456 & 0.601 & 0.455 & 0.845 \\
\textbf{TR}   & 0.604 & 0.572 & 0.702 & 0.924 & 1.665 \\
\textbf{TL}   & 0.597 & 0.560 & 0.706 & 2.796 & 5.717 \\
\textbf{MD}   & 0.666 & 0.629 & 0.779 & 0.385 & 0.792 \\
\textbf{MR}   & 0.733 & 0.701 & 0.830 & 1.121 & 1.967 \\
\textbf{ML}   & \textbf{0.737} & \textbf{0.704} & \textbf{0.834} & 8.283 & 17.744 \\
\bottomrule
\end{tabular}%
}
\caption{Aggregate performance of the static retrieval pipelines. Bold indicates the best static pipeline.}
\label{tab:main_results}
\end{table}

\begin{table}[h]
\centering
\resizebox{\linewidth}{!}{%
\begin{tabular}{c cc cc cc}
\toprule
& \multicolumn{2}{c}{\textbf{RetrievalRouter}}
& \multicolumn{2}{c}{\textbf{Arabzadeh et al.}}
& \multicolumn{2}{c}{\textbf{Oracle}} \\
\cmidrule(lr){2-3}
\cmidrule(lr){4-5}
\cmidrule(lr){6-7}
$\lambda$ &
\textbf{nDCG} & \textbf{Latency} &
\textbf{nDCG} & \textbf{Latency} &
\textbf{nDCG} & \textbf{Latency} \\
\midrule
0.00 & \textbf{0.755}\textsuperscript{$\dagger$} & 0.821 & 0.712 & \textbf{0.522}\textsuperscript{$\dagger$} & 0.901 & 0.927 \\
0.10 & \textbf{0.755}\textsuperscript{$\dagger$} & 0.666 & 0.715 & \textbf{0.476}\textsuperscript{$\dagger$} & 0.901 & 0.340 \\
0.30 & \textbf{0.742}\textsuperscript{$\dagger$} & 0.469 & 0.706 & \textbf{0.382}\textsuperscript{$\dagger$} & 0.901 & 0.335 \\
0.50 & \textbf{0.707}\textsuperscript{$\dagger$} & 0.314 & 0.678 & \textbf{0.276}\textsuperscript{$\dagger$} & 0.893 & 0.306 \\
0.70 & \textbf{0.630} & \textbf{0.148} & 0.624 & 0.171 & 0.829 & 0.197 \\
1.00 & 0.510 & 0.034 & 0.510 & 0.034 & 0.510 & 0.019 \\
\bottomrule
\end{tabular}%
}
\caption{Aggregate performance of the adaptive retrieval pipelines. Bold indicates the better deployable adaptive method. $\dagger$ denotes significance ($p<0.001$).}
\label{tab:adaptive_results}
\end{table}

\FloatBarrier

\subsection{Routing Distribution}
\label{sec:routing_distribution}

\begin{figure}[h]
    \centering
    \includegraphics[width=0.95\linewidth]{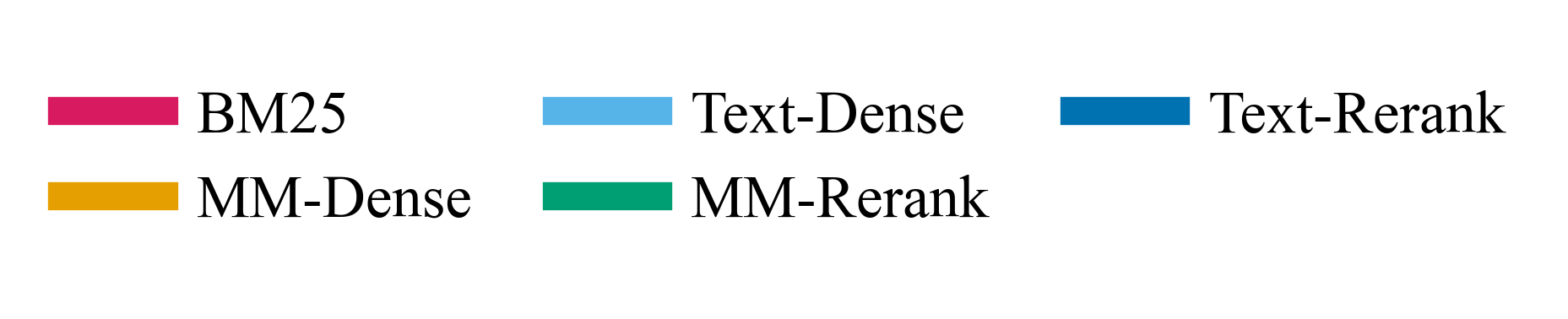}
    \par\vspace{-0.9em}
    
    \begin{subfigure}[t]{0.98\linewidth}
        \centering
        \includegraphics[width=\linewidth]{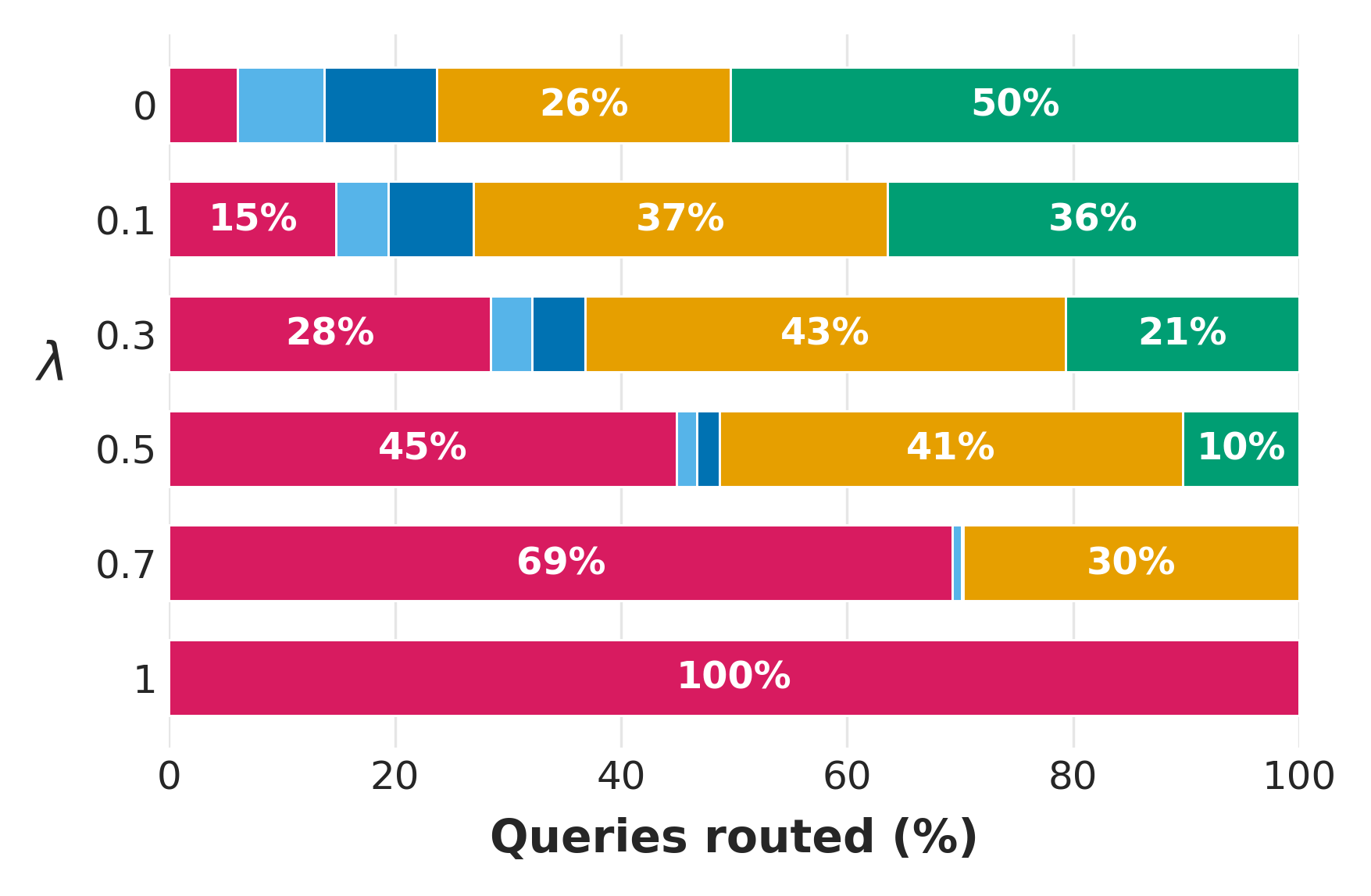}
        \caption{RetrievalRouter}
        \label{fig:routing_distribution_router}
    \end{subfigure}

    \begin{subfigure}[t]{0.98\linewidth}
        \centering
        \includegraphics[width=\linewidth]{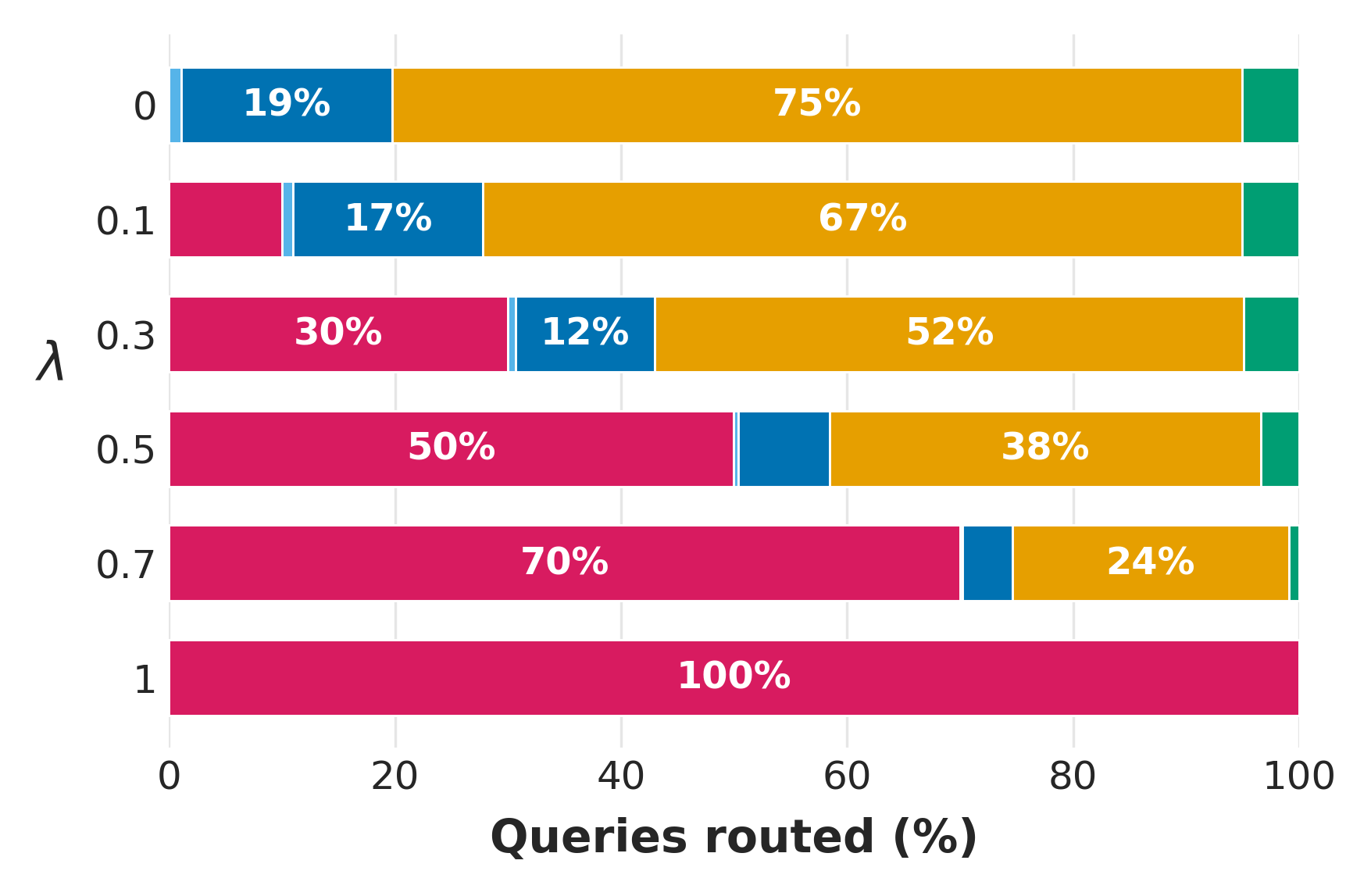}
        \caption{Arabzadeh et al. (2021)}
        \label{fig:routing_distribution_baseline}
    \end{subfigure}

    \caption{Pipeline-selection distributions of adaptive methods across the accuracy--efficiency frontier.}
    \label{fig:routing_distributions}
\end{figure}

\paragraph{Pipeline allocation shifts across the frontier.} To understand how the adaptive strategy-selection baseline of \citet{arabzadeh2021predicting} and RetrievalRouter produce their respective accuracy--efficiency frontiers, we analyze how each method distributes queries across the five selectable retrieval pipelines. Figure~\ref{fig:routing_distributions} compares these routing distributions as the preference shifts from accuracy to efficiency.

\paragraph{Soft targets enable flexible routing.} The difference in routing behavior follows directly from the two training objectives. The strategy-selection baseline assigns each query a single hard label corresponding to the cheapest pipeline that successfully retrieves a relevant page. It therefore cannot become entirely accuracy focused, because its labels reward the first pipeline that succeeds rather than the pipeline that achieves the highest retrieval quality. Even at the equivalent $\lambda=0$ setting, where all emphasis is placed on accuracy, the baseline assigns most queries to MD, the cheapest neural pipeline that succeeds for many queries. This leaves performance on the table when a heavier reranking pipeline could produce a better ranking. RetrievalRouter instead learns from soft targets derived from the complete per-query reward vector. At $\lambda=0$, the latency term vanishes and the targets reflect only retrieval quality, allowing the router to select expensive reranking pipelines whenever their accuracy gains justify them. As $\lambda$ increases, the targets gradually shift toward MD and BM25, enabling RetrievalRouter to adapt its allocation to the desired accuracy--efficiency trade-off.

\paragraph{Cheaper pipelines are not always inferior.} Even at the fully accuracy-focused setting of $\lambda=0$, RetrievalRouter assigns 6.0\% of queries to BM25. Because latency has no influence in this setting, these selections show that the router learns where sparse term matching is effective rather than treating BM25 merely as a low-cost fallback. In contrast, the strategy-selection baseline uses its budget to control how many queries remain on BM25, treating the sparse pipeline as replaceable as more computational budget becomes available. At its equivalent $\lambda=0$ setting, every query is therefore routed to a neural pipeline, preventing BM25 from being selected even when it is the best fit. 

The per-dataset results reinforce this distinction; even at $\lambda=0$, RetrievalRouter routes 20.0\% of Wiki-SS queries to BM25, where BM25 outperforms four of the six neural static pipelines, while the adaptive baseline routes none to BM25 (Appendix~\ref{apx:per_dataset}, Figure~\ref{fig:per_dataset_routing}).

\subsection{Router Decision Quality}

\begin{figure}[h]
    \centering

    \begin{subfigure}[t]{\linewidth}
        \centering
        \includegraphics[width=\linewidth]{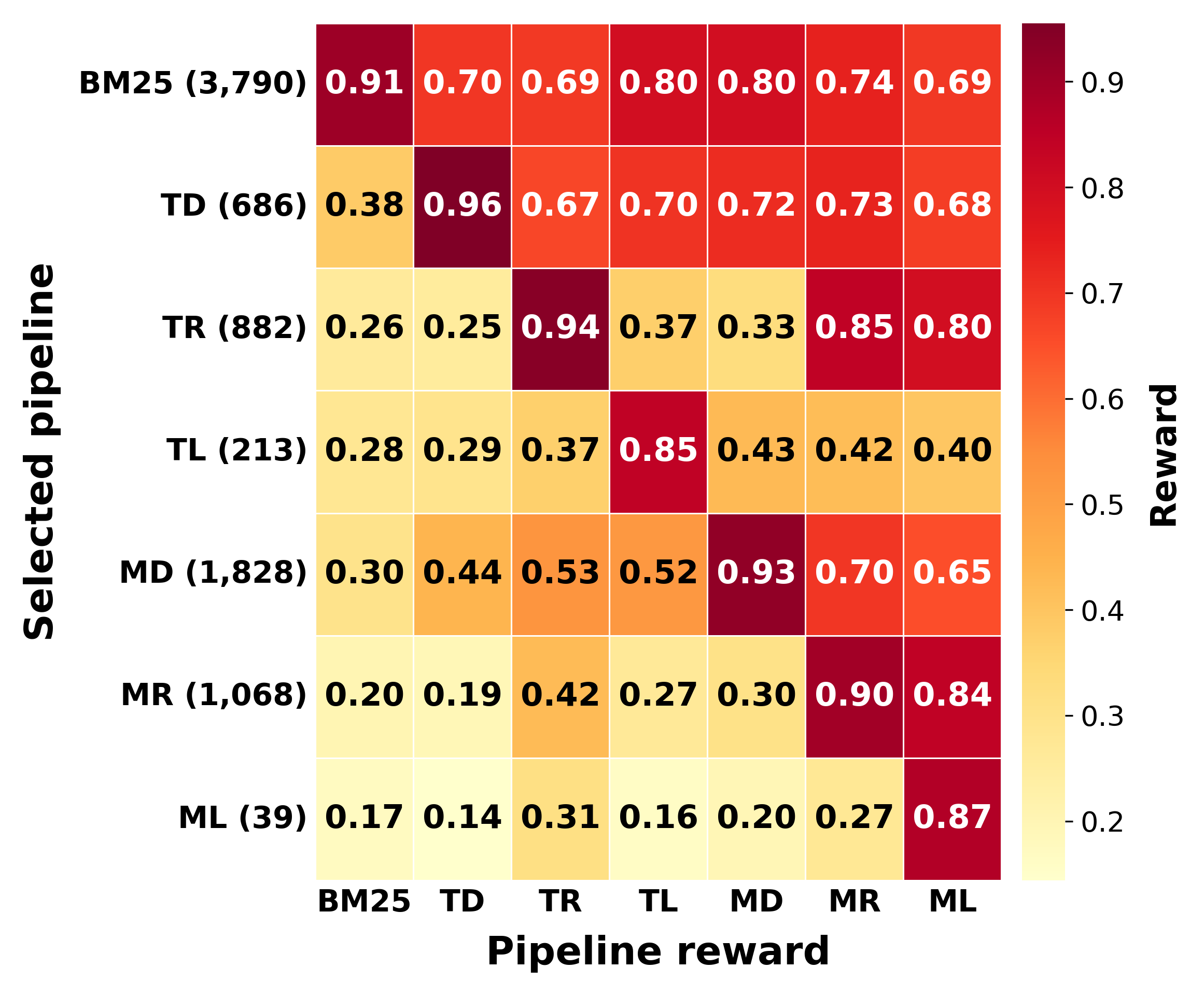}
        \caption{Oracle ($\lambda=0.1$)}
        \label{fig:heatmap_oracle}
    \end{subfigure}

    \begin{subfigure}[t]{\linewidth}
        \centering
        \includegraphics[width=\linewidth]{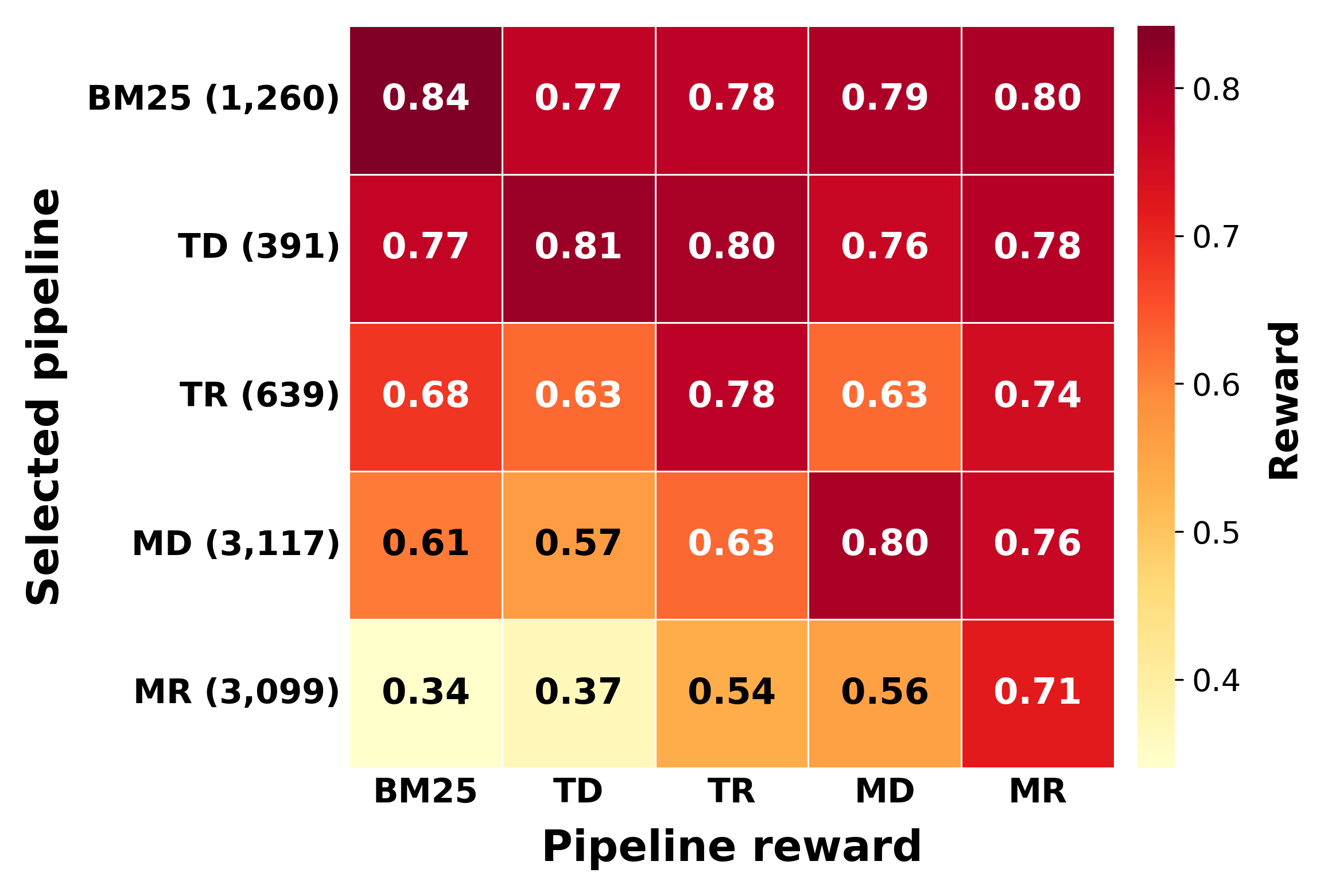}
        \caption{RetrievalRouter ($\lambda=0.1$)}
        \label{fig:heatmap_router}
    \end{subfigure}

    \caption{Reward heatmaps for Oracle and RetrievalRouter decisions. Rows indicate the selected pipeline and columns report the reward of each pipeline.}

    \label{fig:heatmaps}
\end{figure}

To verify that the router learns meaningful selection, we visualize routing decisions as reward heatmaps in Figure~\ref{fig:heatmaps}. Each row corresponds to a selected pipeline, and each column reports the reward it received on the same set of queries. We use $\lambda = 0.1$, since it is closest to the Pareto knee.

\paragraph{Diagonal dominance.} Both the oracle and the router produce strong diagonal patterns: when a pipeline is selected, it is close to optimal for the queries it received. Non-selected pipelines consistently underperform on the same queries. The router does not pick certain pipelines globally; it activates each pipeline selectively for queries where it is the right choice.

\paragraph{Pure late-interaction is rarely chosen.} The Oracle selects TL for 213 queries and ML for 39, compared with 882 for TR and 1,068 for MR. The rerank variants therefore capture most queries where fine-grained interaction is useful, although pure late interaction remains optimal for a very small subset.

\paragraph{Reranking is not always necessary.} Not every query benefits from reranking either; in some cases, adding the rerank pass adds unnecessary cost or even actively degrades retrieval quality. The clearest example is Multimodal-Dense. Its aggregate accuracy across the suite (0.67 nDCG@5) is much lower than its late-interaction counterparts (0.74 nDCG@5), yet on the 1,828 queries where the Oracle selects it, MD achieves an average reward of 0.93. Adding the rerank step (MR) on those same queries drops the reward to 0.70.

\FloatBarrier

\subsection{Modality Sensitivity}
\label{apx:modality}

To understand why no single modality suffices across our benchmark suite, we examine how text and multimodal pipelines respond to visual complexity and textual nuance.  To characterize the visual complexity of each dataset, we compute a per-page visual density score using DocLayout-YOLO \cite{zhao2024doclayout}, defined as the fraction of page area covered by non-textual elements (tables, figures, charts). Scores lie in $[0,1]$, with higher values indicating heavier visual content. 

\begin{figure}[h]
\centering
\includegraphics[width=\linewidth]{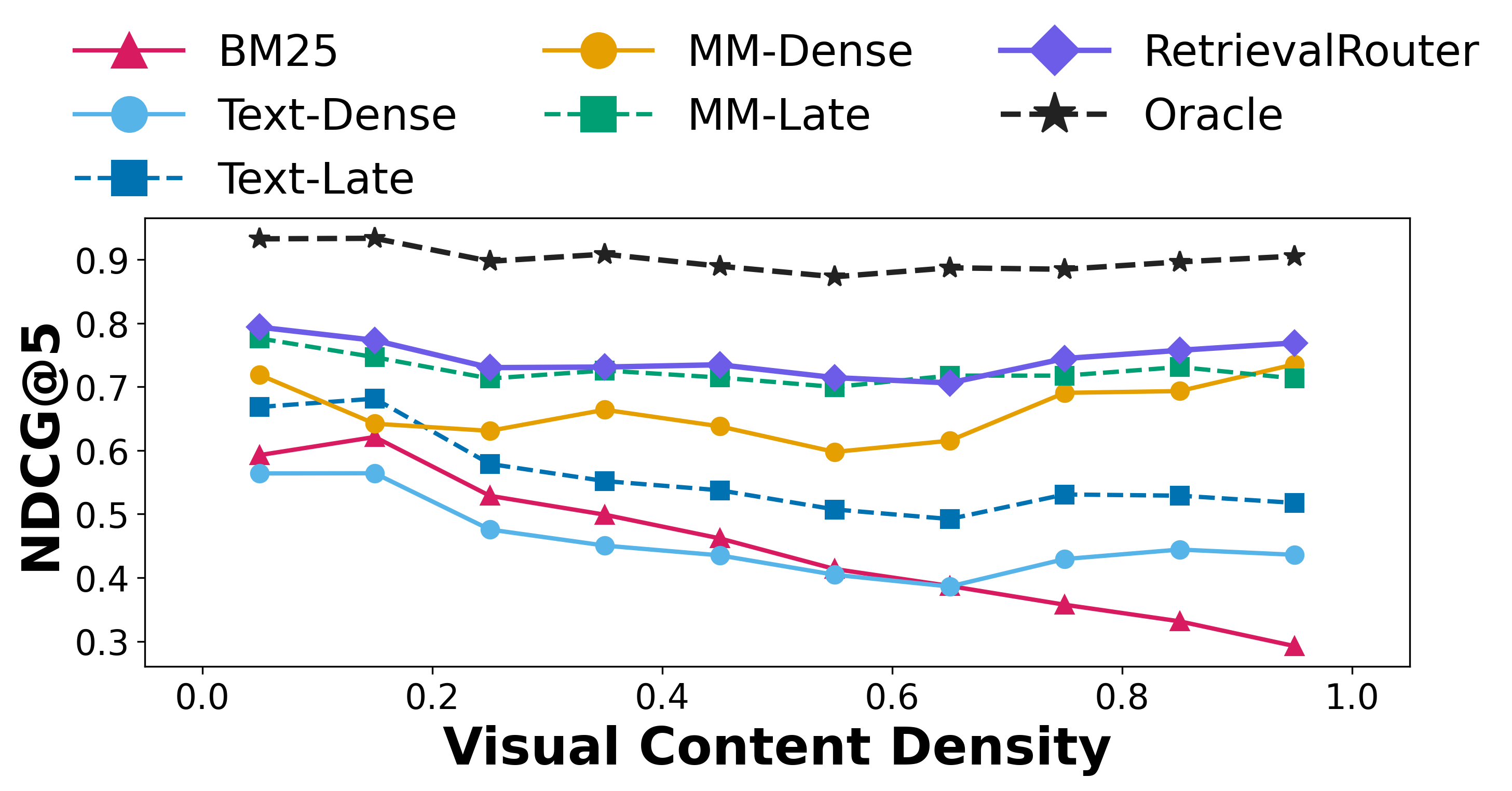}
\caption{nDCG@5 across visual content density.}
\label{fig:visual_density}
\end{figure}

\paragraph{Text pipelines degrade under visual complexity.} Figure~\ref{fig:visual_density} shows that BM25 is the most sensitive to visual content, falling from 0.593 nDCG@5 in the lowest-density bin to 0.292 in the highest. TD and TL also decline as density rises because text extraction linearizes documents and discards the spatial structure needed for tables, charts, and figures. In contrast, the multimodal pipelines remain substantially more stable. 

\begin{figure}[h]
    \centering
    \begin{subfigure}[b]{\linewidth}
        \centering
        \includegraphics[width=\linewidth]{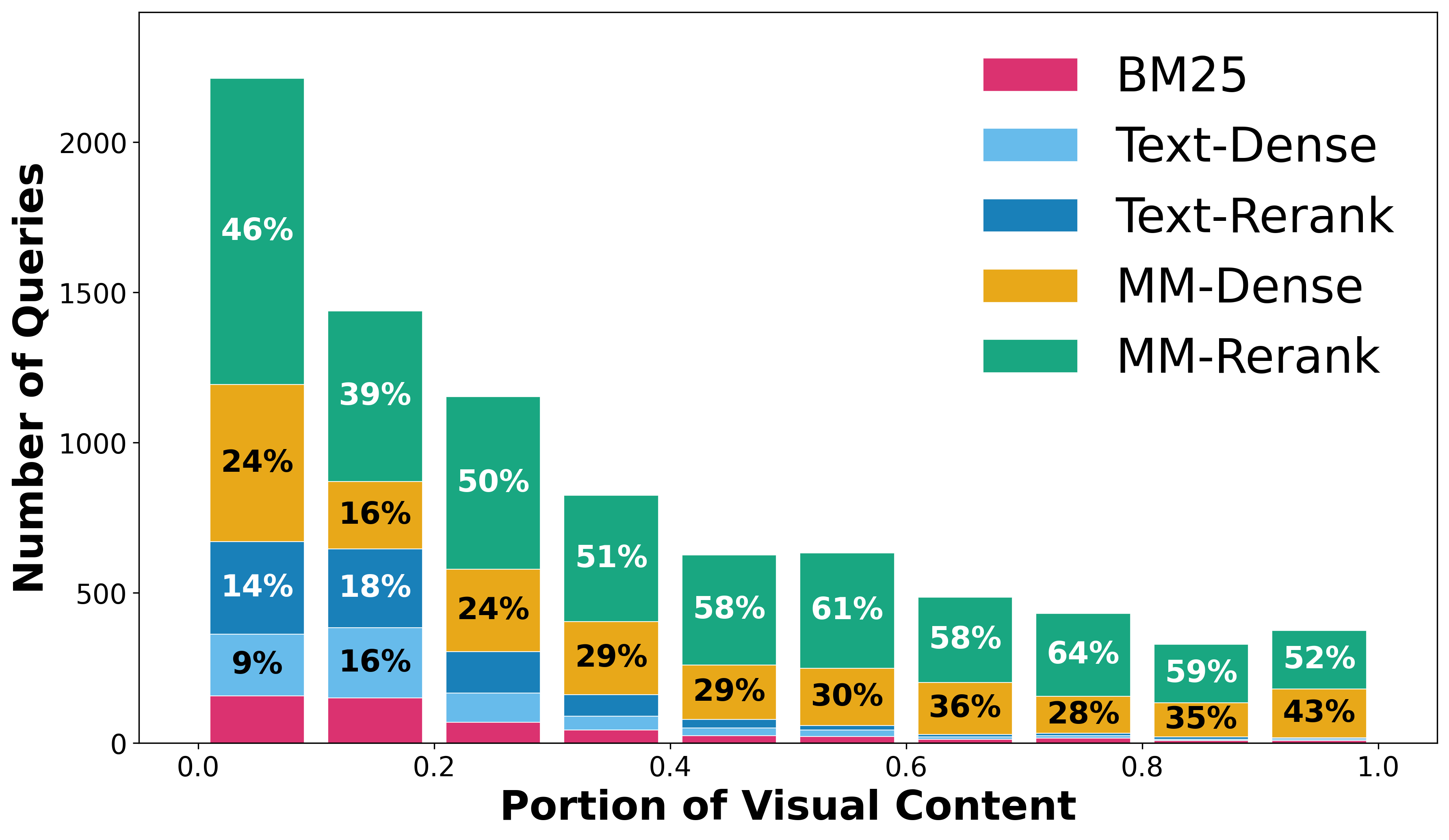}
        \caption{$\lambda=0$}
    \end{subfigure}
    \begin{subfigure}[b]{\linewidth}
        \centering
        \includegraphics[width=\linewidth]{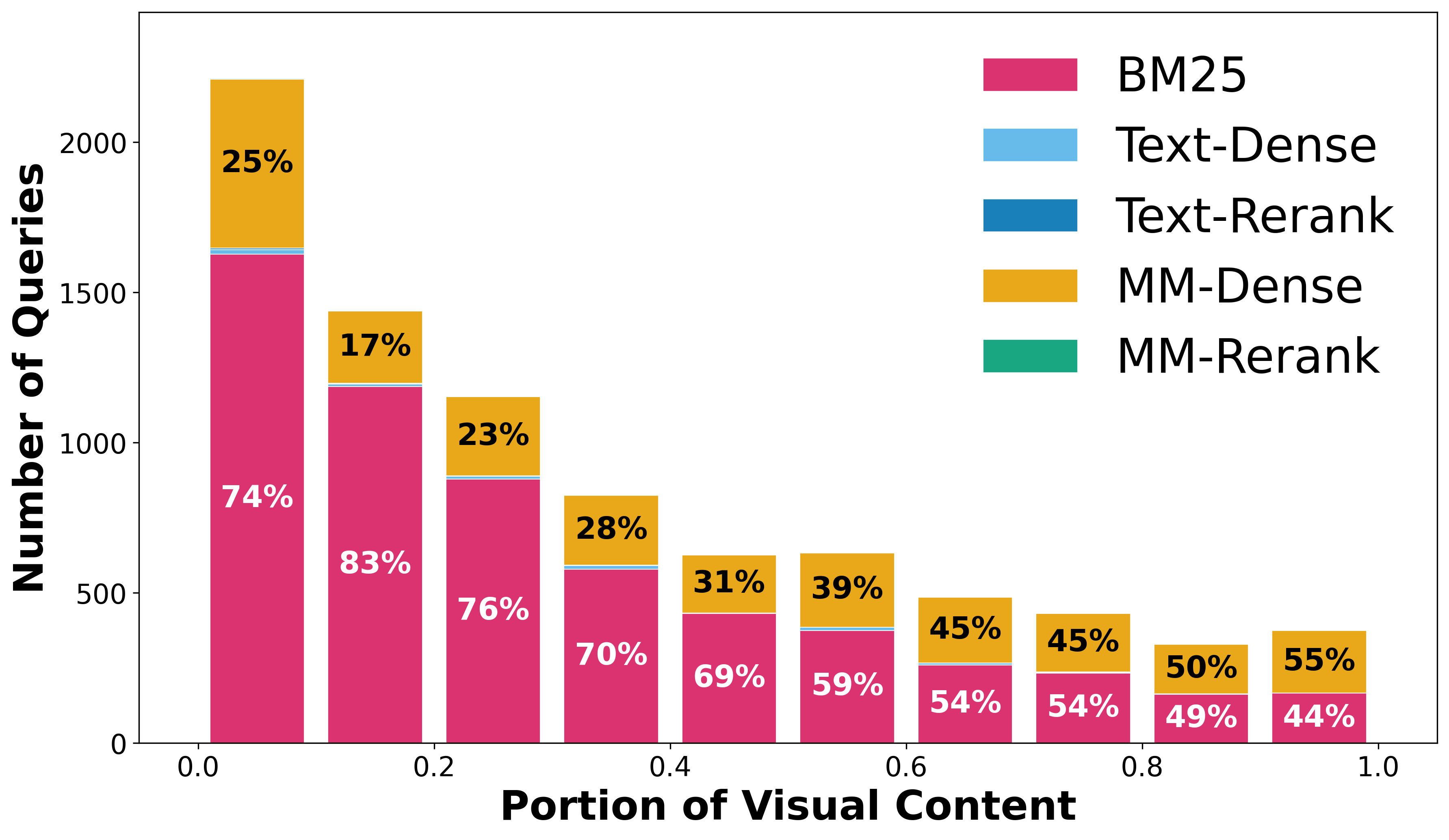}
        \caption{$\lambda=0.7$}
    \end{subfigure}
    \caption{RetrievalRouter pipeline-selection distributions across visual-density bins.}
    \label{fig:pipeline_selection}
\end{figure}

RetrievalRouter tracks this gradient (Figure~\ref{fig:pipeline_selection}): at $\lambda=0$, the share of queries routed to multimodal pipelines rises from 69.7\% to 95.2\% across the density range. Even at $\lambda=0.7$, BM25 usage falls from 73.6\% to 44.3\%, while MD rises from 25.4\% to 55.5\%. Visual complexity influences routing even when the objective prioritizes efficiency.

\paragraph{Multimodal pipelines struggle with textual nuance.} The reverse failure also exists. Table~\ref{tab:per_dataset_results} shows that multimodal pipelines do not dominate every dataset. On Wiki-SS, TL reaches 0.784 nDCG@5, outperforming both ML (0.743) and MR (0.732). BM25 also reaches 0.745, exceeding both multimodal late-interaction variants. Wiki-SS emphasizes long-range textual reasoning and fine-grained semantic integration, a regime in which visual encoders' patch-level structure can become a liability. RetrievalRouter recognizes this structure and reaches 0.785 nDCG@5, slightly exceeding the best static pipeline. Multimodal representations are essential for layout-heavy documents, but can dilute the sequential textual structure that language-intensive tasks depend on.

\FloatBarrier
\section{Conclusions}
\label{sec:conclusions}

In this work, we showed that the accuracy--latency trade-off in document retrieval is a routing problem, not an architectural one. RetrievalRouter formalizes this view: it dispatches each query to the cheapest pipeline that can answer it, reserving expensive pipelines for queries that genuinely require them. The practical consequence is that practitioners no longer face a binary choice between an accurate pipeline that is too slow to deploy and a fast pipeline that fails on the documents in their workload. 

Our contributions are threefold: (i) a systematic empirical analysis of modern retrieval pipelines along the modality and architecture axes across 11 benchmarks; (ii) RetrievalRouter itself, the first system to perform joint modality--architecture routing over a single underlying corpus; and (iii) a query-level pipeline-selection benchmark releasing per-query best-pipeline labels across more than 80{,}000 queries.

Our analysis of static pipelines surfaces several findings. (i) Late-interaction pipelines outperform their dense counterparts across the suite, but (ii) their rerank variants recover all of that accuracy essentially at a fraction of the cost, making reranking the best choice for deployment of late-interaction. (iii) Reranking is nonetheless not always beneficial: it adds unnecessary latency when the dense retriever already answers the query correctly, and in some cases actively degrades performance. On the modality axis, (iv) text-based pipelines degrade under high visual complexity, while (v) multimodal pipelines degrade under textual nuance. Finally, (vi) lexical retrieval remains competitive on datasets where exact keyword matching is effective, making BM25 a viable accuracy-oriented choice rather than merely a low-cost fallback.

RetrievalRouter recognizes and avoids these failure modes. It dominates every neural static configuration across both accuracy and latency: 2.5\% higher nDCG@5 than the strongest static baseline while being $12.4\times$ faster, and 3.0\% higher than the deployment-standard rerank variant at $1.7\times$ lower latency. Furthermore, compared with prior adaptive strategy selection methods, RetrievalRouter achieves significantly higher nDCG@5 in accuracy-oriented settings, while matching or numerically outperforming them in both effectiveness and latency in latency-oriented settings.

\newpage
\clearpage

\section*{Limitations}
\label{sec:limitations}

\textbf{The Storage Cost.} A primary criticism of multi-model routing involves operational storage costs. Unlike a static pipeline that maintains a single index, RetrievalRouter maintains four vector indices and a separate BM25 lexical index. Table~\ref{tab:storage_cost} reports the vector index sizes (excluding lexical BM25). The storage footprint is dominated by the multimodal-late (ColPali) index, which holds multi-vector patch embeddings. For the combined corpus, the multimodal-late index ($\approx 39$~GB) is more than $13\times$ larger than the text-dense index ($\approx 3$~GB). This trade-off is fundamental: our approach uses more storage for reduced inference latency. In cloud environments where storage is cheap and latency directly affects user experience and API costs, caching representations minimizes demands on costly GPU compute time. In contrast, storage-constrained environments may benefit from a static pipeline. Against this backdrop, it is essential to examine alternative approaches, such as cascading systems, and compare them with routing.

\begin{table}[htbp]
    \centering
    \small
    \resizebox{\linewidth}{!}{%
    \begin{tabular}{lrr}
        \toprule
        \textbf{Pipeline} & \textbf{Index Size (MB)} & \textbf{Relative Factor} \\
        \midrule
        Text-Dense       & 3,003  & $1.0\times$ \\
        Text-Late        & 8,308  & $2.8\times$ \\
        Multimodal-Dense & 2,618  & $0.9\times$ \\
        Multimodal-Late  & 39,023 & $13.0\times$ \\
        \bottomrule
    \end{tabular}%
    }
    \caption{Vector-index size comparison.}
    \label{tab:storage_cost}
\end{table}

\textbf{The VRAM Cost.} Similarly, another key trade-off in our approach is the increased GPU usage required to achieve lower latency. Maintaining all pipelines at once requires substantially more VRAM (40~GB) than standard approaches. Consequently, our approach is best suited for \textit{latency-critical} applications where hardware costs are secondary to user experience. 

\textbf{Cross-Domain Generalization.} Our experimental setup uses an intra-dataset split (80/10/10), evaluating in-domain generalization where query distributions are known at deployment time. A potential limitation is that the router may learn domain-specific lexical cues (for instance, financial vocabulary as a proxy for high visual density) rather than the underlying semantic structure. In practice, enterprise document search engines operate over fixed corpora with known query domains at compile time, making in-domain splits representative of production scenarios. Nonetheless, evaluating zero-shot cross-domain generalization, such as testing on multimodal visual document benchmarks like ViDoRe \cite{vidore2026}, remains a valuable direction for future research.

\textbf{Query-Only Semantic Ambiguity.} A key limitation of RetrievalRouter is its reliance on the query text alone. While the router outperforms all static baselines, a substantial gap remains relative to a relevance-aware Oracle because the query text alone can be semantically ambiguous. For example, queries like ``summarize the table on page 5'' are linguistically identical regardless of whether the referenced document contains a visual infographic or extracted text. Purely semantic routing is thus underdetermined when the optimal pipeline depends on the target document's latent layout rather than query intent. Overcoming this ceiling requires active probing, such as exploratory dense retrieval or partial metadata inspection, to resolve document-level structural cues before full pipeline dispatch. Finally, our routing action space focuses on bi-encoder and late-interaction architectures; expanding the policy to decide when expensive generative LLM-based rerankers (e.g., listwise or pointwise reasoning) or downstream generator model sizes are justified represents a compelling direction for compound AI systems.

\section*{Ethical Considerations}

We foresee no major ethical concerns or potential risks in our work. All retrieval models, encoders, and benchmark datasets used in this study are publicly released and open-sourced, and are used under their original licenses. We introduce no new human-subject data and conduct no human evaluation.

\paragraph{Data and Reproducibility.}
The 11 benchmarks span financial filings, scientific papers, and open-domain web sources. We release our code, training scripts, and per-query oracle labels across more than 80{,}000 queries to support reproducibility and further work on adaptive retrieval. Our code and data are available at \url{https://github.com/emrekuruu/retrieval-router}

\paragraph{Bias and Generalization.}
The training distribution is restricted to English-language documents from financial, scientific, and open-domain corpora. Practitioners deploying the router outside these distributions should evaluate performance on representative in-domain queries.

\bibliography{bibtex} 

\newpage
\clearpage

\appendix

\section{Experimental Details}
\label{apx:experimental_details}

\subsection*{Dataset Statistics}

Table~\ref{tab:datasets} reports the query count, corpus size, and average document length for each dataset. 

\begin{table}[h!]
\centering
\small
\resizebox{\columnwidth}{!}{
\begin{tabular}{llrrc}
\toprule
\textbf{Benchmark} & \textbf{Subset} & \textbf{\# Queries} & \textbf{\# Docs} & \textbf{Avg. Tokens} \\
\midrule
\multirow{2}{*}{\textbf{REAL-MM-RAG}} & FinReport & 853 & 2,687 & 1,053 \\
 & FinSlides & 1,048 & 2,280 & 275 \\
\midrule
\multirow{4}{*}{\textbf{T2-RAGBench}} & FinQA & 6,232 & 2,789 & 965 \\
 & ConvFinQA & 3,431 & 1,806 & 966 \\
 & VQAnBD & 9,772 & 1,787 & 780 \\
 & TAT-DQA & 27,127 & 2,758 & 852 \\
\midrule
\multirow{5}{*}{\textbf{MMDocRAG}} & ArxivQA & 9,034 & 4,749 & 1,110 \\
 & Wiki-SS & 14,968 & 12,752 & 777 \\
 & MP-DocVQA & 5,581 & 2,350 & 388 \\
 & SciQAG & 4,496 & 2,595 & 1,196 \\
 & DUDE & 2,561 & 2,073 & 516 \\
\bottomrule
\end{tabular}
}
\caption{Dataset statistics. The suite covers over 80,000 queries across financial, scientific, and open domains.}
\label{tab:datasets}
\end{table}

\subsection*{Implementation Details}

\paragraph{Data split.} Each source dataset is independently split into 80\% train, 10\% validation, and 10\% test. The test set is held out during oracle labeling and router training.

\paragraph{Oracle generation.} For each training query, all seven pipelines are executed once to record nDCG@5 and latency. While the base nDCG and latency metrics are collected once via the oracle, the composite reward and soft targets are dynamically recomputed for each specific $\lambda$ objective during training using Equations~\ref{eq:reward} and \ref{eq:soft_target}.

\paragraph{Hyperparameters.} The decision head is a single linear projection from the 1024-dimensional query representation to the five routing logits, with dropout 0.1 applied to the pooled representation. The LoRA adapters use rank 16, $\alpha = 32$, and dropout 0.05, introducing approximately 4M trainable parameters out of Qwen3-0.6B's 600M total. Queries are tokenized to a maximum of 128 tokens. We train for 2 epochs using AdamW with learning rate $1 \times 10^{-4}$,  batch size 16 with gradient accumulation of 2 (effective batch size 32), weight decay 0.01, and linear warmup over the first 10\% of steps followed by cosine decay, on a single NVIDIA H100 in bfloat16 precision. Training takes approximately 20 minutes per $\lambda$ setting. LoRA and optimizer hyperparameters follow standard practice for LoRA fine-tuning; the temperature $\tau = 0.1$ was selected on the validation split. Finally, all models are trained with a random seed of 42, and the reported test-set numbers are from a single training run per $\lambda$ setting.

\paragraph{Inference Cost.} On an NVIDIA H100, the router adds approximately 15ms of inference latency per query. For perspective, the Text-Dense pipeline runs in roughly 400ms end-to-end, while Multimodal-Late runs in roughly 8s. The routing overhead therefore corresponds to approximately 3.8\% of TD latency and under 0.2\% of ML latency.

\section{Per-Dataset Analysis}
\label{apx:per_dataset}

\begin{figure}[h]
    \centering
    \includegraphics[width=0.99\linewidth]{figures/routing_distribution_legend.png}
    \par\vspace{-0.8em}

    \begin{subfigure}[t]{0.99\linewidth}
        \centering
        \includegraphics[width=\linewidth]{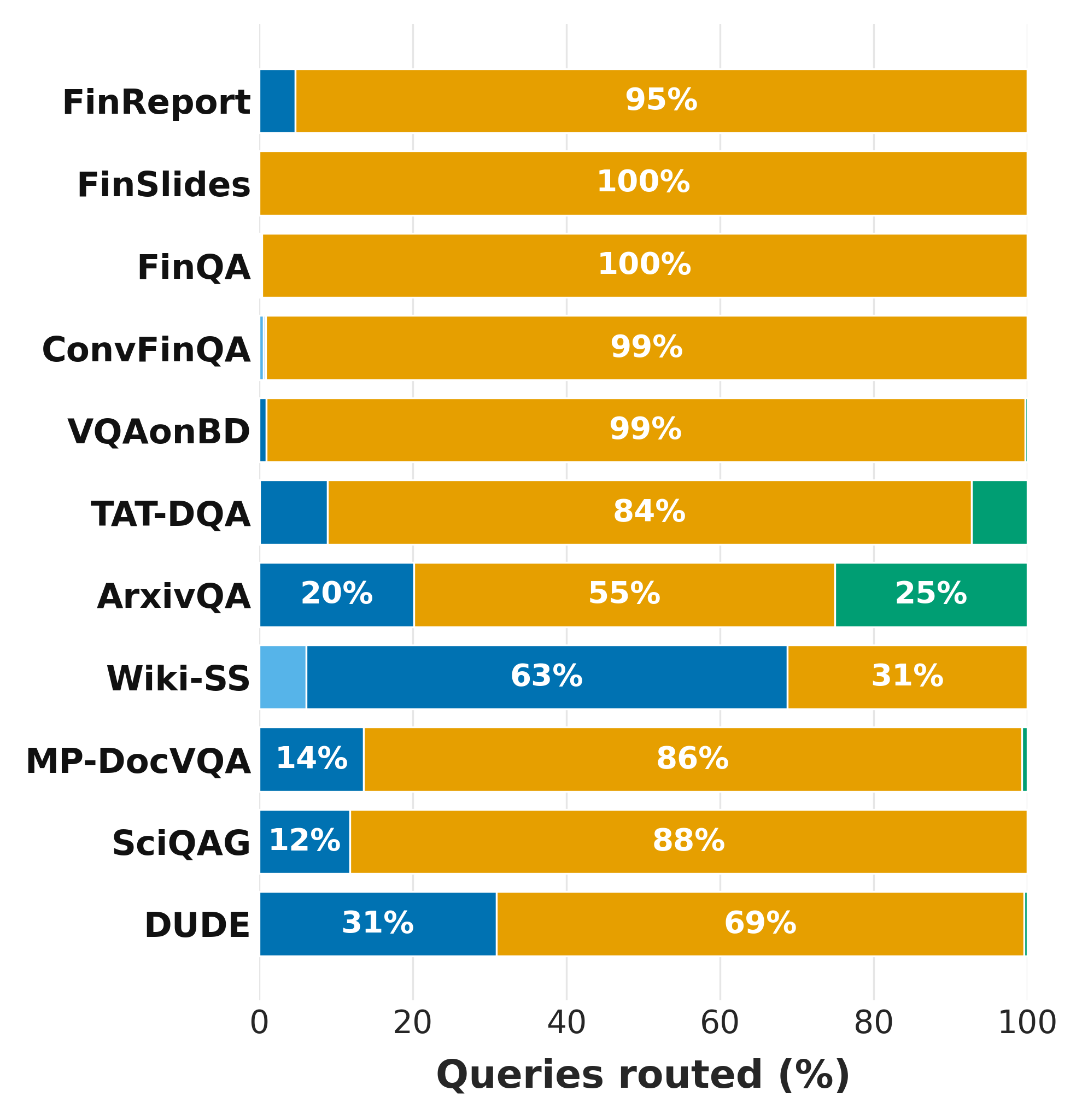}
        \caption{Arabzadeh et al. (2021)}
    \end{subfigure}

    \begin{subfigure}[t]{0.99\linewidth}
        \centering
        \includegraphics[width=\linewidth]{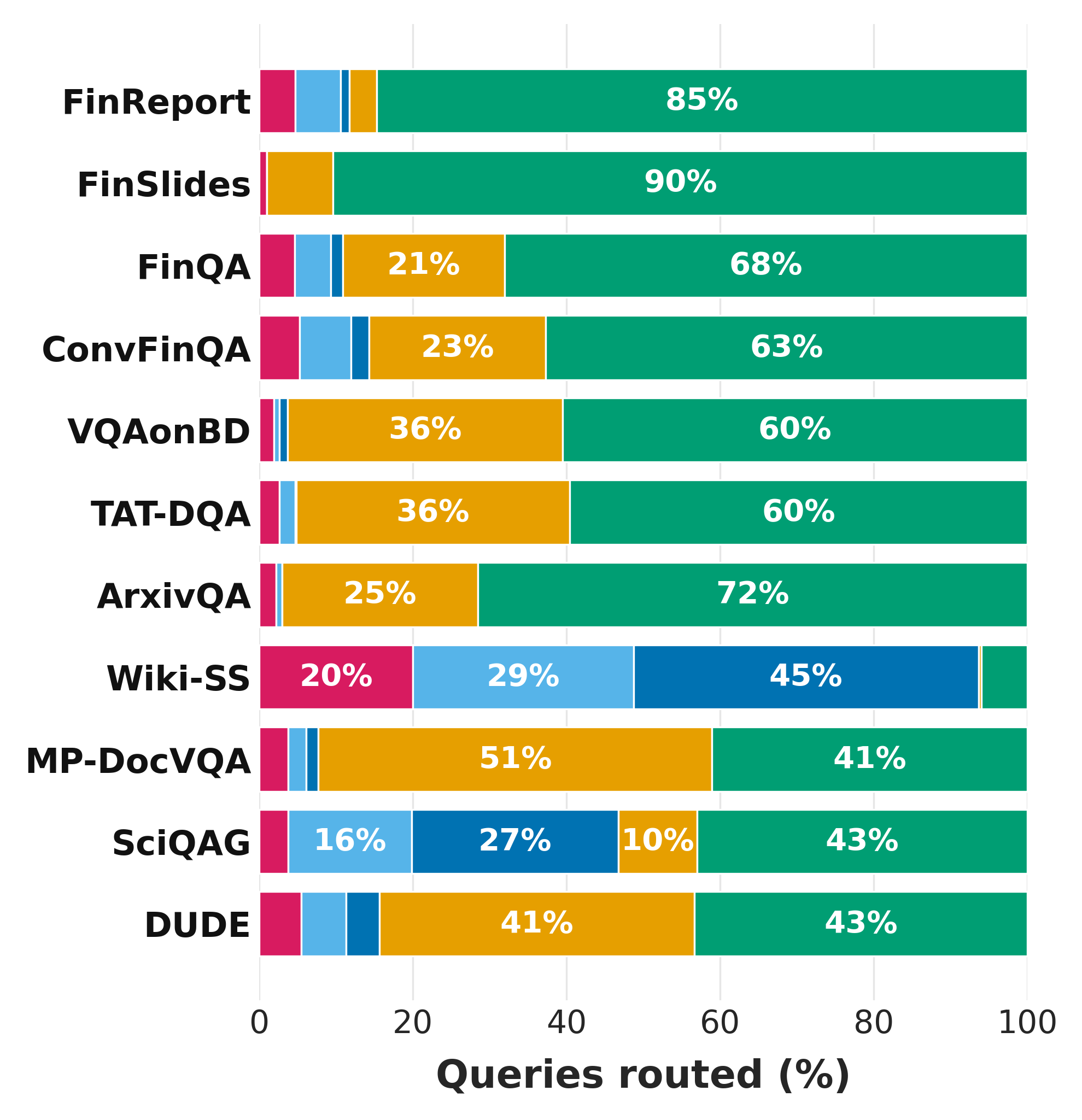}
        \caption{RetrievalRouter}
    \end{subfigure}

    \caption{Per-dataset routing distributions at $\lambda=0$.}
    \label{fig:per_dataset_routing}
\end{figure}

Table~\ref{tab:per_dataset_results} reports per-dataset nDCG@5, MRR@5, and Recall@5 for the static pipelines, adaptive methods, and Oracle. Figure~\ref{fig:per_dataset_routing} compares the pipeline allocations of RetrievalRouter and Arabzadeh et al. at $\lambda=0$, showing how the two training objectives respond to dataset-specific retrieval demands.

From our results, we observe that RetrievalRouter allocates computation based on where each pipeline improves retrieval rather than following a fixed cost ordering. Arabzadeh et al. routes 84--100\% of queries from the six financial datasets to MD because its hard-label objective favors the cheapest pipeline that succeeds. However, MR is the strongest static pipeline on all six datasets. RetrievalRouter recognizes this difference and routes 60--90\% of their queries to MR instead (Table~\ref{tab:per_dataset_ndcg}). The allocation changes when a dataset favors another modality. On Wiki-SS, where text pipelines outperform multimodal pipelines, RetrievalRouter sends 94\% of queries to the text pipelines. 

The results also show that RetrievalRouter learns when cheaper pipelines are appropriate. Lexical matching is particularly effective on Wiki-SS, where BM25 outperforms four of the six neural static pipelines. RetrievalRouter routes 20.0\% of Wiki-SS queries to BM25 even at $\lambda=0$, when latency has no influence on the reward. This shows that BM25 is selected for its retrieval effectiveness rather than merely as a low-cost fallback. On MP-DocVQA and DUDE, RetrievalRouter instead divides most queries between MD and MR, reflecting query-level variation within the same dataset. This dataset-aware allocation translates into broader gains: RetrievalRouter exceeds the strongest static pipeline on five datasets for nDCG@5 and MRR@5 and on four for Recall@5, while Arabzadeh et al. does so on only two datasets for each metric.

\begin{table*}[t]
\centering
\small
\setlength{\tabcolsep}{3.8pt}
\renewcommand{\arraystretch}{0.92}
\begin{subtable}[t]{\textwidth}
\centering
\resizebox{\textwidth}{!}{%
\begin{tabular}{lccccccc|cc|c}
\toprule
& \multicolumn{7}{c}{\textbf{Static Pipelines}} & \multicolumn{2}{c}{\textbf{Adaptive Routers}} & \textbf{Upper Bound} \\
\cmidrule(lr){2-8}\cmidrule(lr){9-10}\cmidrule(lr){11-11}
\textbf{Dataset} & \textbf{BM25} & \textbf{TD} & \textbf{TL} & \textbf{TR} & \textbf{MD} & \textbf{ML} & \textbf{MR} & \textbf{Arab.} & \textbf{Ours} & \textbf{Oracle} \\
\midrule
FinReport & 0.456 & 0.527 & 0.726 & 0.736 & 0.621 & 0.803 & \textbf{0.807} & 0.620 & 0.798 & 0.938 \\
FinSlides & 0.183 & 0.410 & 0.387 & 0.382 & 0.799 & 0.902 & \textbf{0.903} & 0.799 & 0.886 & 0.949 \\
FinQA & 0.685 & 0.707 & 0.746 & 0.785 & 0.831 & 0.865 & \textbf{0.868} & 0.831 & 0.865 & 0.961 \\
ConvFinQA & 0.663 & 0.716 & 0.782 & 0.808 & 0.815 & 0.862 & \textbf{0.863} & 0.815 & 0.863 & 0.950 \\
VQAonBD & 0.595 & 0.564 & 0.626 & 0.731 & 0.815 & 0.846 & \textbf{0.847} & 0.816 & \underline{0.856} & 0.948 \\
TAT-DQA & 0.361 & 0.331 & 0.446 & 0.428 & 0.610 & 0.652 & \textbf{0.656} & 0.633 & \underline{0.678} & 0.842 \\
ArxivQA & 0.307 & 0.251 & 0.399 & 0.393 & 0.424 & \textbf{0.664} & 0.630 & 0.566 & 0.634 & 0.801 \\
Wiki-SS & 0.745 & 0.702 & \textbf{0.784} & 0.774 & 0.648 & 0.743 & 0.732 & 0.761 & \underline{0.785} & 0.968 \\
MP-DocVQA & 0.441 & 0.423 & 0.604 & 0.628 & \textbf{0.728} & 0.724 & 0.726 & \underline{0.740} & \underline{0.781} & 0.906 \\
SciQAG & 0.722 & 0.731 & 0.871 & 0.879 & 0.804 & 0.881 & \textbf{0.881} & 0.815 & 0.859 & 0.994 \\
DUDE & 0.400 & 0.490 & 0.606 & 0.547 & 0.628 & 0.644 & \textbf{0.647} & \underline{0.717} & \underline{0.730} & 0.901 \\
\bottomrule
\end{tabular}%
}
\caption{nDCG@5}
\label{tab:per_dataset_ndcg}
\end{subtable}
\par\vspace{1.2em}\noindent
\begin{subtable}[t]{\textwidth}
\centering
\resizebox{\textwidth}{!}{%
\begin{tabular}{lccccccc|cc|c}
\toprule
& \multicolumn{7}{c}{\textbf{Static Pipelines}} & \multicolumn{2}{c}{\textbf{Adaptive Routers}} & \textbf{Upper Bound} \\
\cmidrule(lr){2-8}\cmidrule(lr){9-10}\cmidrule(lr){11-11}
\textbf{Dataset} & \textbf{BM25} & \textbf{TD} & \textbf{TL} & \textbf{TR} & \textbf{MD} & \textbf{ML} & \textbf{MR} & \textbf{Arab.} & \textbf{Ours} & \textbf{Oracle} \\
\midrule
FinReport & 0.420 & 0.484 & 0.682 & 0.690 & 0.574 & 0.764 & \textbf{0.770} & 0.576 & 0.758 & 0.921 \\
FinSlides & 0.170 & 0.374 & 0.365 & 0.358 & 0.763 & 0.876 & \textbf{0.876} & 0.763 & 0.857 & 0.932 \\
FinQA & 0.644 & 0.652 & 0.702 & 0.744 & 0.792 & 0.833 & \textbf{0.835} & 0.792 & 0.833 & 0.948 \\
ConvFinQA & 0.618 & 0.669 & 0.744 & 0.774 & 0.781 & 0.835 & \textbf{0.836} & 0.781 & 0.836 & 0.938 \\
VQAonBD & 0.549 & 0.514 & 0.575 & 0.686 & 0.774 & 0.810 & \textbf{0.812} & 0.775 & \underline{0.822} & 0.936 \\
TAT-DQA & 0.328 & 0.299 & 0.408 & 0.394 & 0.567 & 0.613 & \textbf{0.617} & 0.591 & \underline{0.637} & 0.812 \\
ArxivQA & 0.286 & 0.227 & 0.373 & 0.370 & 0.393 & \textbf{0.639} & 0.608 & 0.534 & 0.606 & 0.779 \\
Wiki-SS & 0.710 & 0.672 & \textbf{0.751} & 0.743 & 0.617 & 0.712 & 0.701 & 0.732 & \underline{0.757} & 0.959 \\
MP-DocVQA & 0.413 & 0.388 & 0.573 & 0.606 & 0.697 & 0.699 & \textbf{0.700} & \underline{0.707} & \underline{0.751} & 0.899 \\
SciQAG & 0.679 & 0.685 & 0.841 & 0.850 & 0.766 & 0.855 & \textbf{0.855} & 0.776 & 0.827 & 0.992 \\
DUDE & 0.379 & 0.465 & 0.586 & 0.529 & 0.600 & 0.614 & \textbf{0.617} & \underline{0.695} & \underline{0.700} & 0.889 \\
\bottomrule
\end{tabular}%
}
\caption{MRR@5}
\label{tab:per_dataset_mrr}
\end{subtable}
\par\vspace{1.2em}\noindent
\begin{subtable}[t]{\textwidth}
\centering
\resizebox{\textwidth}{!}{%
\begin{tabular}{lccccccc|cc|c}
\toprule
& \multicolumn{7}{c}{\textbf{Static Pipelines}} & \multicolumn{2}{c}{\textbf{Adaptive Routers}} & \textbf{Upper Bound} \\
\cmidrule(lr){2-8}\cmidrule(lr){9-10}\cmidrule(lr){11-11}
\textbf{Dataset} & \textbf{BM25} & \textbf{TD} & \textbf{TL} & \textbf{TR} & \textbf{MD} & \textbf{ML} & \textbf{MR} & \textbf{Arab.} & \textbf{Ours} & \textbf{Oracle} \\
\midrule
FinReport & 0.565 & 0.659 & 0.859 & 0.871 & 0.765 & \textbf{0.918} & \textbf{0.918} & 0.753 & 0.918 & 0.988 \\
FinSlides & 0.221 & 0.519 & 0.452 & 0.452 & 0.904 & \textbf{0.981} & \textbf{0.981} & 0.904 & 0.971 & 1.000 \\
FinQA & 0.806 & 0.870 & 0.878 & 0.909 & 0.945 & 0.960 & \textbf{0.965} & 0.945 & 0.961 & 0.997 \\
ConvFinQA & 0.796 & 0.857 & 0.895 & 0.907 & 0.913 & 0.942 & \textbf{0.945} & 0.913 & 0.942 & 0.985 \\
VQAonBD & 0.735 & 0.715 & 0.778 & 0.865 & 0.934 & \textbf{0.950} & \textbf{0.950} & 0.936 & \underline{0.956} & 0.985 \\
TAT-DQA & 0.461 & 0.427 & 0.561 & 0.529 & 0.740 & 0.767 & \textbf{0.773} & 0.762 & \underline{0.803} & 0.934 \\
ArxivQA & 0.371 & 0.326 & 0.476 & 0.461 & 0.515 & \textbf{0.735} & 0.694 & 0.660 & 0.717 & 0.868 \\
Wiki-SS & 0.848 & 0.792 & \textbf{0.885} & 0.864 & 0.743 & 0.836 & 0.822 & 0.848 & 0.867 & 0.992 \\
MP-DocVQA & 0.525 & 0.537 & 0.704 & 0.705 & \textbf{0.840} & 0.809 & 0.813 & \underline{0.847} & \underline{0.876} & 0.947 \\
SciQAG & 0.849 & 0.866 & 0.962 & \textbf{0.964} & 0.918 & 0.958 & 0.958 & 0.931 & 0.955 & 1.000 \\
DUDE & 0.466 & 0.569 & 0.664 & 0.607 & 0.718 & 0.736 & \textbf{0.740} & \underline{0.792} & \underline{0.823} & 0.939 \\
\bottomrule
\end{tabular}%
}
\caption{Recall@5}
\label{tab:per_dataset_recall}
\end{subtable}
\caption{Per-dataset effectiveness at $\lambda=0$. Bold marks the best static pipeline; underline marks an adaptive router that strictly outperforms the best static pipeline. Oracle values are excluded from highlighting.}
\label{tab:per_dataset_results}
\end{table*}

\FloatBarrier

\end{document}